\documentclass[10pt,twocolumn,
superscriptaddress,aip,author-year]{revtex4-1}
\usepackage{times}
\usepackage{color}
\usepackage{amsmath}
\usepackage{amssymb}
\usepackage{graphicx}
\renewcommand{\cite}[1]{ \citep{#1} }

\begin{document}

\title{\begin{center}\Large Lateral Gene Transfer from the Dead \end{center} \vskip 1 cm }

\author{Gergely J. Sz\"oll\H{o}si}
\affiliation{Laboratoire de Biom\'etrie et Biologie Evolutive, Centre National de la Recherche Scientifique, Unit\'e Mixte de Recherche 5558, Universit\'e Lyon 1, F-69622 Villeurbanne, France;}
\affiliation{Universit\'e de Lyon, F-69000 Lyon, France;}
\author{Eric Tannier}
\affiliation{Laboratoire de Biom\'etrie et Biologie Evolutive, Centre National de la Recherche Scientifique, Unit\'e Mixte de Recherche 5558, Universit\'e Lyon 1, F-69622 Villeurbanne, France;}
\affiliation{Universit\'e de Lyon, F-69000 Lyon, France;}
\affiliation{Institut National de Recherche en Informatique et en Automatique Rh\^one-Alpes, F-38334 Montbonnot, France}
\author{Nicolas Lartillot}
\affiliation{Centre Robert-Cedergren pour la Bioinformatique,
D\'epartement de Biochimie, Universit\'e de Montr\'eal, Qu\'ebec, Canada}
\affiliation{Laboratoire d'Informatique, de Robotique et de Micro\'electronique de
Montpellier, UMR 5506, CNRS-Universit\'e de Montpellier 2, Montpellier,
France}
\author{Vincent Daubin}
\email[]{vincent.daubin@univ-lyon1.fr}
\affiliation{Laboratoire de Biom\'etrie et Biologie Evolutive, Centre National de la Recherche Scientifique, Unit\'e Mixte de Recherche 5558, Universit\'e Lyon 1, F-69622 Villeurbanne, France;}
\affiliation{Universit\'e de Lyon, F-69000 Lyon, France;}

\begin{abstract}
In phylogenetic studies, the evolution of molecular sequences is assumed to have taken place along the phylogeny traced by the ancestors of extant species. In the presence of lateral gene transfer (LGT), however, this may not be the case, because the species lineage from which a gene was transferred may have gone extinct or not have been sampled. Because it is not feasible to specify or reconstruct the complete phylogeny of all species, we {\color{black} must} describe the evolution of genes outside the represented phylogeny by modelling the speciation dynamics that gave rise to the complete phylogeny. We demonstrate that if the number of sampled species is small compared to the total number of existing species, the overwhelming majority of gene transfers involve speciation to, and evolution along extinct or unsampled lineages. We show that the evolution of genes along extinct or unsampled lineages can to good approximation be treated as those of independently evolving lineages described by a few global parameters. Using this result, we derive an algorithm to calculate the probability of a gene tree and recover the maximum likelihood reconciliation given the phylogeny of the sampled species. Examining 473 near universal gene families from 36 cyanobacteria, we find that nearly a third of transfer events -- $28\%$ -- {\color{black}appear to have} topological signatures of evolution along extinct species, but only approximately $6\%$ of transfers trace their ancestry to before the common ancestor of the sampled cyanobacteria. 
\end{abstract}
\maketitle

\begin{quote}
\emph{``From the first growth of the tree, many a limb and branch has decayed and dropped off; and these lost branches of various sizes may represent those whole orders, families, and genera which have now no living representatives, and
which are known to us only from having been found in a fossil state.''}
\\Charles Darwin, \emph{On the Origins of Species}. London, 1859
\end{quote}
Most of the diversity of life that ever existed on earth has gone extinct and can only be glimpsed from the fossil record. Although the comparative approach allows the reconstruction of some morphological and genetical characteristics of ancestral species, it is only informative for species that have founded extant lineages. Yet, the information enclosed in genome sequences is abundant and particularly meaningful for the reconstruction of the descent and evolution of their carriers\cite{zp,Boussau:2010cr,David:2011zr}, so much so that it may have recorded accounts of extinct lineages. This possibility exists because the success of lateral gene transfer (LGT) as an evolutionary process implies that each gene possesses its own, unique history, which is not necessarily confined to the history of those species that have survived\cite{Maddison:1997ly,Galtier:2008zr,Fournier:2009nx,Abby:2012uq}.

Several models have recently been developed to reconcile seemingly contradictory gene phylogenies with the species phylogeny by tracing the path on the species phylogeny along which they evolved as result of a series of speciations, gene duplications, LGT and losses\cite{tofigh_2009,doyon_2011,David:2011zr,Szollosi:2012fk,Szollosi:2012fkk}. None of these models, however, take into consideration the fact that, in the presence of LGT, gene trees record evolutionary paths along the complete species tree, including extinct and unsampled branches, and not only along the phylogeny of the species in which they reside today. This is the case because, as first noted by Maddison\cite{Maddison:1997ly} and later elaborated by Gogarten et al.\ \cite{Zhaxybayeva:2004fk,Fournier:2009nx}, while LGT events imply that the donor and receiver lineages existed at the same time, the donor lineage might have subsequently become extinct, or more generally, might not have been sampled. 

Here we demonstrate that, if the number of species considered in the species phylogeny is small compared to the total number of species, the overwhelming majority of gene transfers involve speciation to, and evolution along extinct or unsampled species. Furthermore, we show that, if this condition is met, the evolution of genes along the unrepresented parts of the species phylogeny can to good approximation be treated as those of independently evolving lineages, the behaviour of which depends only on the global parameters of the speciation dynamics. This in turn allows us to derive the probability of observing a gene phylogeny by extending the ODT model introduced previously \cite{Szollosi:2012fkk}. Applying our model to a dataset derived from 36 cyanobacterial species, we {\color{black}perform a preliminary} assessment of the phylogenetic signal for the evolution of transferred genes along extinct species. 

\section*{A minimal model of speciation and gene birth and death}

\begin{figure}
\begin{center}
\centerline{\includegraphics[width=1.\columnwidth]{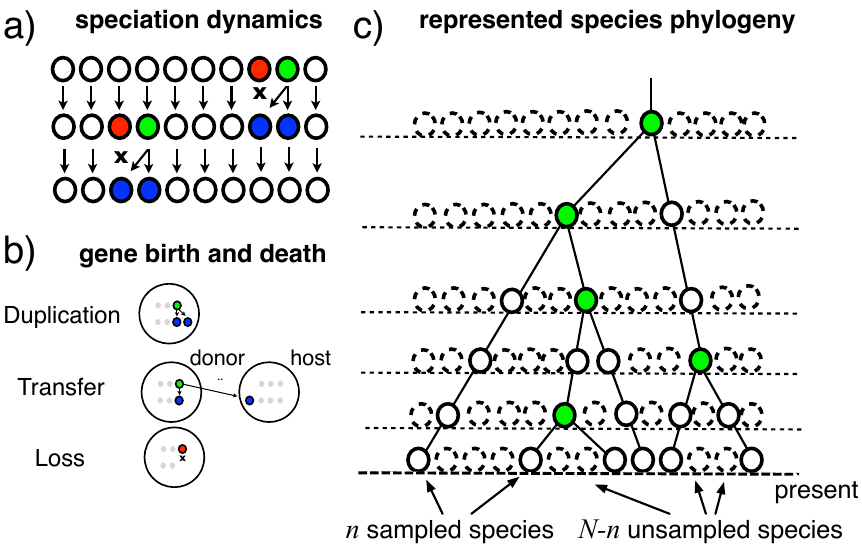}}
\caption{{\bf Gene trees are the result of the combination of speciation and gene birth and death.} As a minimal description we consider: a) that for each of the $N$ species at a rate $\sigma$, a speciation occurs, during which the species is succeeded by two descendants, and a random species suffers extinction; b) at a rate $\delta$ per gene, a gene duplicates, i.e., it is succeeded by two gene copies in the same genome, at a rate $\tau/(N-1)$ per gene per host species, a gene is transferred, resulting in one copy each in the donor and host species, and finally with a rate $\lambda$ per gene, a gene is lost. The represented phylogeny c) corresponds to the tree spanned by the $n$ sampled species. A branch of the represented tree corresponds to a series of speciation events, but only the last of these, the speciation event that gives rise to two represented lineages (filled circles, green {\color{black}online}) is explicitly present for internal branches as the speciation node terminating the branch. The number of unrepresented species (dashed circles) is always much larger than the number of represented species (full circles). \label{model}}
\end{center}
\end{figure}

It is not feasible to specify, much less to reconstruct, the complete phylogeny of all species that ever existed. To describe the evolution of genes outside the represented phylogeny -- along lineages that have become extinct or whose descendants have not been sampled -- we must resort to modelling the speciation dynamics that gave rise to the complete phylogeny. Modelling the dynamics of speciation provides a stochastic model of the evolution of unrepresented lineages that can be used to describe gene histories given knowledge of the represented phylogeny and a few global parameters. 

As a minimal model of speciation, here, we assume that the number of species $N$ is constant, and that the dynamics of speciation is modeled by a continuous time Moran process\cite{Moran:1962cr}. That is, for each species at rate $\sigma$, a speciation occurs during which the species gives rise to two descendants and a randomly chosen species goes extinct (cf.\ Fig.\ref{model}a). The central assumption we make is that, of the $N$ species existing at present (i.e., $t=0$), we sample only a small fraction $n \ll N$. In general, the validity of this assumption depends on the phylogenetic problem considered, but should almost always be met for major groups of bacteria and archaea, where the number of species that potentially exchange genes by LGT is inevitably much larger than the number of sampled species, even in large scale studies \cite{Ochman:2000dq,Torsvik:2002fk}.

To describe the evolution of genes within the genomes of species we assume genes to evolve independently according to a birth-and-death process that consists of gene duplication, transfer and loss\cite{tofigh_2009,Szollosi:2012fk,Szollosi:2012fkk}. As shown in Fig.\ref{model}b, a gene in the genome of any of the $N$ species can: i) be duplicated at rate $\delta$; ii) be transferred from a donor species to any of the other $N-1$ possible host species at a rate $\tau/(N-1)$; or iii) be lost at a rate $\lambda$. Genes copies can also be born and be lost as a result of the speciation dynamics: iv) at the species level lineages experience speciation at a rate $\sigma$, in which case they are replaced by two copies in the two new species, or v) suffer extinction at an identical rate $\sigma$. A branch $e$ of the represented tree $S$ in general corresponds to a series of speciation events, however, as shown in Fig.\ref{model}c, only the last one of these, the speciation event that gave rise to two represented lineages, is explicitly present for internal branches as the (green {\color{black}online}) speciation node terminating the branch.

\section*{Almost all transfers involve speciation}

To understand what fraction of transfers involves evolution along unrepresented species we must compare the relative rate of transfers that are \emph{direct transfers }between branches of the represented phylogeny $S$, {\color{black} and} \emph{indirect transfers} that result in a gene returning to $S$ after exiting it via speciation or transfer to unrepresented species. 

 To compare the contribution of indirect transfers and direct transfers to observed gene histories, we consider first only direct transfers and indirect transfers that involve a speciation to an unrepresented species. To describe the shape of the species tree generated by the Moran process introduced above, we can use the coalescent approach. Here, under Kingman's coalescent, the time to the most recent common ancestor of the $n$ sampled species is of the order of $2N/\sigma (1-1/n) \approx 2N / \sigma$\cite{kingman_1982}. This implies that the expected number of unrepresented speciation events per branch of the species tree is much larger than one, being of the order of $ \sigma \times 2N / \sigma/(2n-2) \approx N/n \gg 1$, as there are $(2n-2)$ branches of $S$. This suggests that for any pair of coexisting branches of the represented tree, a gene that descends from one of the branches and is transferred to the other, is likely to have experienced a speciation event "away" from the represented phylogeny spanned by the $n$ sampled species before being transferred back to it. 

To quantify the above argument we can compare the expected number of transfers from branch $f$ to branch $e$ of the represented phylogeny, resulting from either a direct transfer or a more complex history involving a speciation event. Clearly, if the branches do not overlap in time, the expected number of direct transfers is zero. To consider overlapping branches let us consider for simplicity that both $e$ and $f$ are terminal branches - similar results can be derived for any other pair of overlapping branches.
The expected branch lengths are then $E(t_e)=E(t_f)\approx N / \sigma n $, with overlap $\min(t_e,t_f)\lesssim N / \sigma n$. {\color{black}Integrating over possible transfer times,} the expected number of direct transfers is then 
\begin{align} 
T_{\mathrm{direct}} &\lesssim \int_{0}^{ \frac{N}{ n \sigma } } \frac{\tau}{N-1} \mathrm{d}t_e' = \frac{\tau}{2}\frac{1}{n} \frac{1}{(N-1)} \left[ \frac{2 N } { \sigma } \right]. \label{direct_eq}
\end{align}

\begin{figure}
\begin{center}
\centerline{\includegraphics[width=1.\columnwidth]{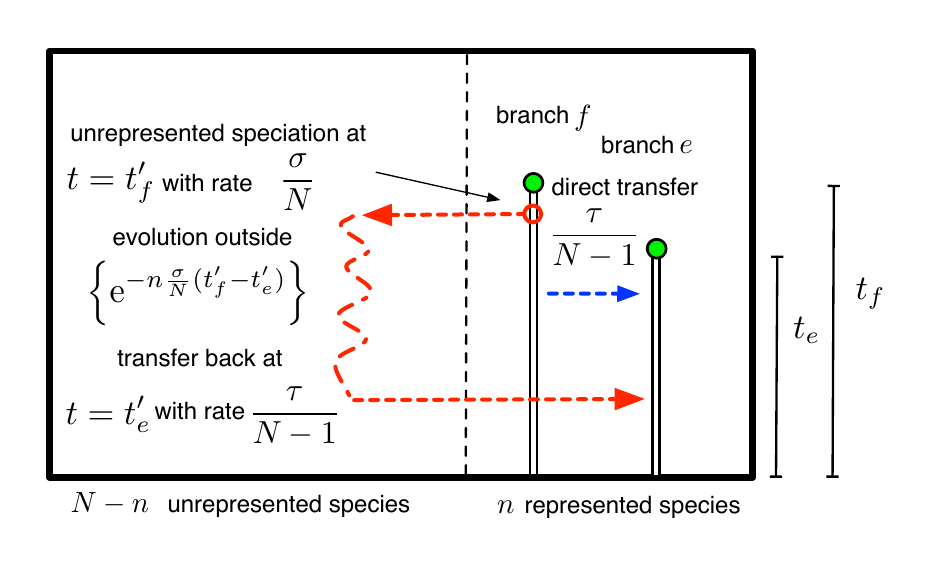}}
\caption{{\bf The overwhelming majority of transfers involve evolution along unrepresented species.} A direct transfers (dark grey, blue {\color{black}online}) between two terminal branches of the represented phylogeny {\color{black} occurs with  rate $\frac{\tau}{N-1}$ and} involves a single transfer event. An indirect transfer (light grey, red {\color{black}online}) that leaves an indistinguishable record in the gene tree topology. {\color{black} To count indirect transfers, we trace their history backwards in time: transfer back to the host branch on the represented tree (branch $e$) occur with a rate $\frac{\tau}{N-1}$ from each of the $N-n$ unrepresented species, of these we are only concerned with ones which descend from the relevant donor branch (branch $f$), the number of these can be calculated using the exponential coalescence probability and the rate of unrepresented speciations $\frac{\sigma}{N}$ from the donor branch (branch $f$).}} 
  \label{direct}
\end{center}
\end{figure}

To estimate the expected number of indirect transfers that are topologically indistinguishable from the above direct transfers we can reason backwards in time as illustrated in Fig.\ref{direct}: i) the rate at which a transfer occurs from each of the $(N-n)$ unrepresented species to branch $e$ is $\tau/(N-1)$; ii) the probability of this gene lineage \emph{not} coalescing back to any of the $n$ branches of the represented tree during a time interval $t$ is $\exp(-n \sigma / N t )$, and iii) the rate at which it coalesces with branch $f$ is $\sigma/N$. Integrating over possible speciation {\color{black} and transfer} times gives:
\begin{align} 
T_{\mathrm{indirect}}&\simeq \int_{0}^{ \frac{N}{ n \sigma } }\! \int_{t_e'}^{ \frac{N}{ n \sigma } }\! \frac{(N-n) \tau}{N-1} \left\{ \mathrm{e}^{- n\frac{ \sigma }{N} (t_f'-t_e') } \right\} \frac{\sigma}{N} \ \mathrm{d}t_f' \mathrm{d}t_e' \nonumber\\
&= \frac{ \tau }{2} \frac{1}{n^2} \frac{ (N-n)}{ \mathrm{e} (N-1) } \left[ \frac{ 2 N } { \sigma } \right]. \label{indirect_eq}
\end{align}

Equations (\ref{direct_eq}) and (\ref{indirect_eq}) show that if the number of sampled species is small compared to total number of species ($n\ll N$), then the expected number of direct transfers is small compared to indistinguishable indirect ones ($T_{\mathrm{direct}} \ll T_{\mathrm{indirect}}$), i.e., \emph{the contribution of direct transfers to observed gene histories is negligible}. 

To compare the two types of possible indirect transfers back to $S$ -- those exiting via speciation and those via transfer - we must contrast the rate $\sigma$ at which gene copies exit branch $f$ as a result of speciation and the rate $ \tau /(N-1) \times (N -1 -n ) \approx \tau$ that gene copies exit as a result of transfer. Estimates of $\tau$, and more generally gene birth and death rates, are available from several sources, all of which agree that the expected number of gene birth and death events per branch is below unity. Models that consider the dynamics of the number of homologous gene copies along a species phylogeny (referred to as phylogenetic profiles) \cite{Csuros:2009zr} have consistently found that birth and death rate is of the same order, with an excess of loss compensated by origination of new families, in agreement with phenomenological models of gene family size distribution \cite{Karev:2002dq,Szollosi:2012fk}. In a detailed study, Cs\H{u}r\"os et al.\ found for $28$ archaea that the expected number of birth events (duplication and gain) is $0.12$ and that the expected number of losses is $0.36$ \cite{Csuros:2009zr} per branch per gene. More recently, the ODT model that attempts to explicitly explain the evolution of multi-copy gene trees (representative of complete genomes) along an ultrametric species tree has arrived at similar results\cite{Szollosi:2012fkk}, finding for 36 cyanobacterial genomes $\delta \approx \tau \approx 0.2,\ \lambda \approx 1$, in units corresponding to a tree with unit height. Assuming, as above, that the time to the most recent common ancestor of the sampled species is of the order $2N/\sigma$, i.e., the expected number of gene copies (per gene) exiting a branch of $S$ is proportional to $N/n$, while the number exiting as a result of transfer is less than one. Since the rate at which a gene that has exited the represented phylogeny returns to $S$ as a results of transfer at some point in the future is independent of the mode of exit from $S$, we can conclude that \emph{indirect transfers are dominated by paths that include a speciation}. 

In summary, if the number of sampled species is small compared to the total number of species, transfers in observed gene histories are dominated by paths that include a speciation to an unrepresented species and subsequent transfer back to the represented tree.

\section*{The probability of observing a gene tree}

\begin{figure*}
\begin{center}
\centerline{\includegraphics[width=2.\columnwidth]{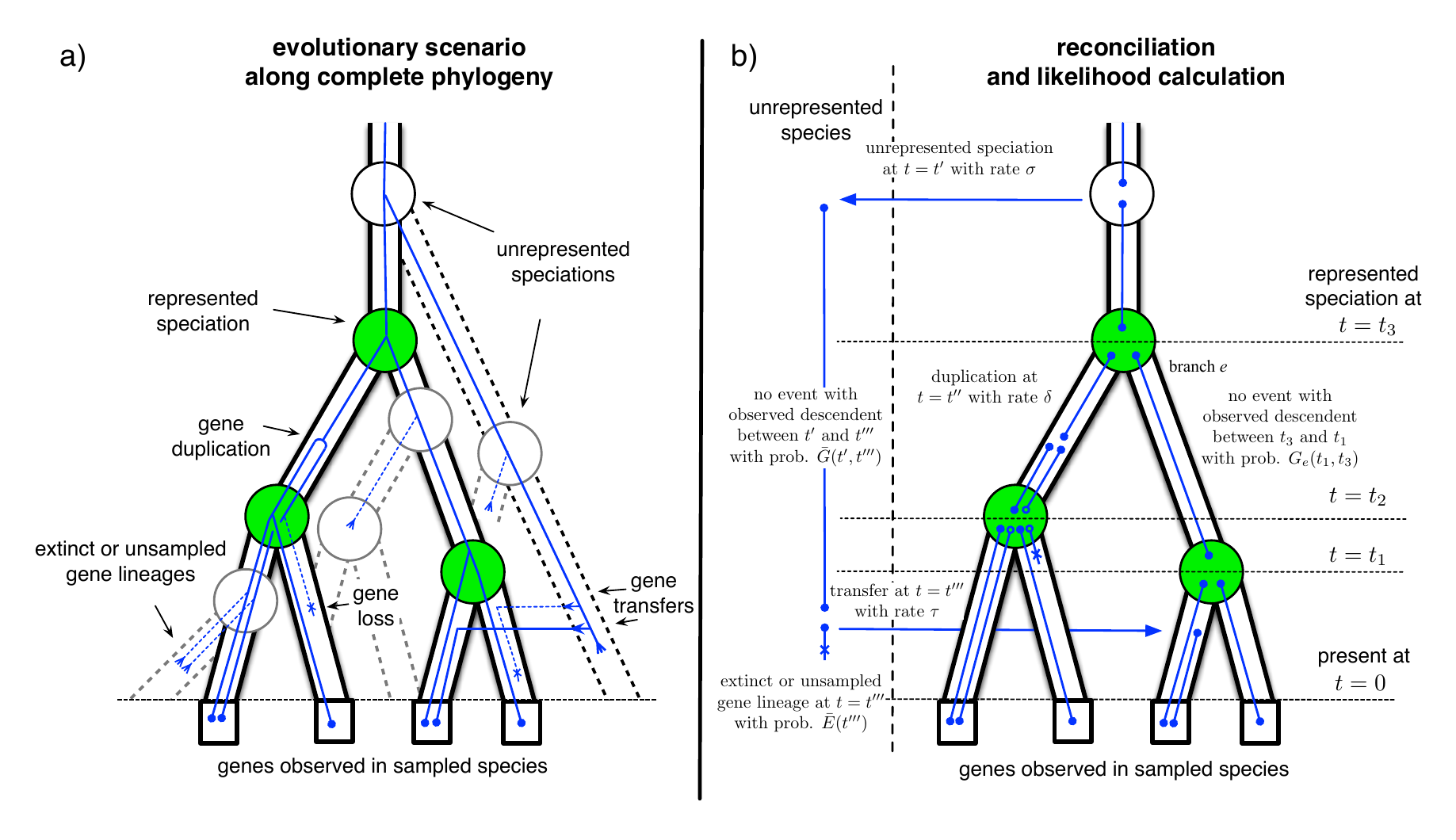}}
\caption{\color{black}  { \color{black} \bf Reconciling gene trees with the complete phylogeny. }\label{rec} a) shows an evolutionary scenario that involves a transfer event from an unrepresented species. The represented phylogeny is shown as a solid tube with filled circles (green {\color{black}online}) corresponding to represented speciations. The unrepresented phylogeny is indicated by dashed tubes, with white circles corresponding to unrepresented speciations (cf.\ Fig.\ref{model}c). The continuous line traces the gene tree spanned by genes in sampled species that is the result of a series of birth and death events along the complete phylogeny. b) a reconciliation of the gene phylogeny from (a), corresponding to the evolutionary scenario depicted in (a). In general we do not know the evolutionary scenario that has generated the gene phylogeny. However, we can use the dynamic programming algorithm described in the text to calculate the likelihood of the gene tree by summing over all possible reconciliations, i.e., all ways to draw the gene tree into the species using speciation, duplication, transfer and loss events (cf.\ Eqs.\ref{Pe}-\ref{Pe0} and  Fig.\ref{diag}) in the Appendix. The likelihood calculation uses the rate of different events ($\sigma,\delta,\tau$ and $\lambda$) together with functions describing the extinction ($E_e$ and $\bar E$) and the propagation ($G_e$ and $\bar G$) of gene linages (cf.\ Eqs.\ref{ee}-\ref{barg}).  }  
\end{center}
\end{figure*}

Reconciling gene trees with the species tree requires iterating over possible paths along which a gene tree may have been generated by a series of speciations, duplications, transfers and losses {\color{black}(Fig.\ref{rec})}. In existing methods \cite{tofigh_2009,doyon_2011,Szollosi:2012fk,Szollosi:2012fkk}, this is accomplished by only considering paths along the represented phylogeny and using a dynamic programming approach exploiting the independence of gene birth and death events, and by extension gene lineages. 

While gene duplication, transfer and loss can {\color{black} reasonably} be modeled as independent birth and death events, speciation and extinction necessarily involve the simultaneous birth and death of many genes. Along the represented phylogeny, speciation events are fully specified and can be explicitly taken into account \cite{Szollosi:2012fkk}. This is not the case, however, for speciation and extinction events that occur in the unrepresented part of the phylogeny, or do not correspond to speciation nodes of the represented phylogeny. Therefore, unrepresented speciations result in non-independence of gene lineages.

Consider for instance the probability $\bar E_k(t)$ that $k$ genes present at time $t$ in a species not ancestral to the sample of $n$ extant species leave no observed descendant. Conditional on the complete phylogeny, $\phi$ including all extinct species lineages, gene lineages are independent, and therefore $\bar E_k(t|\phi) = \{\bar E(t|\phi)\}^k$. Averaging over all complete phylogenies compatible with the phylogeny reconstructed based on the $n$ species, however, results in $\langle \bar E_k(t|\phi) \rangle = \langle \{\bar E(t|\phi)\}^k \rangle \neq \langle \bar E_k(t|\phi) \rangle^k$, which is not a product of $k$ independent factors.

On the other hand, $n \ll N $ implies that $\bar E(t) \approx 1$. Introducing the notation $\bar E(t|\phi)=1-\epsilon(t|\phi)$ and $\bar E(t)=1-\epsilon(t)$, and neglecting second and higher order terms in $\epsilon(t|\phi)$ and $\epsilon(t)$ {\color{black} we have}:
\begin{align}
 \bar E_k(t) &= \big \langle \bar E_k(t|\phi) \big \rangle_\phi =\big \langle \left\{\bar E(t|\phi)\right\}^k \big\rangle_\phi = \big \langle \{1- \epsilon(t|\phi)\}^k \big \rangle_\phi \nonumber \\ 
&\simeq \big \langle \{1- k \epsilon(t|\phi)\}\big \rangle_\phi = \{1- k \big \langle\epsilon(t|\phi)\big\rangle_\phi \} = 1- k \epsilon(t) \nonumber \\ &\simeq \left\{ 1- \epsilon(t) \right\}^k= \left\{\bar E(t)\right\}^k. \label{indep}
\end{align}
A similar argument can be derived for $k$-gene propagator $\bar G_k(s,t)$ (see Appendix). Therefore, if $n \ll N $, then to good approximation, the evolution of two genes observed in the same unrepresented species can be treated as independent without specifying the full phylogeny. 

Under the above assumption that unrepresented speciation and extinction events can be considered in a gene-wise independent manner, we can describe the evolution of gene copies that appear as single gene lineages when observed from the present. {\color{black} We can calculate: i) the extinction probability $E_e(t)$ that a gene seen at time $t$ on branch $e$ of $S$ leaves no \emph{observed descendant} i.e., no descendant exists at time $t=0$ in the genome of any of the $n$ sampled species; ii) the extinction probability $\bar E(t)$ that a gene seen at time $t$ in an unrepresented species leaves no observed descendant; iii) the single gene propagation probabilities $G_e(s,t)$ that all observed descendants of a gene seen at time $s$ on branch $e$ descend from a descendant seen at a later time $t<s$ on branch $e$; and iv) $\bar G(s,t)$ the probability that all observed descendants of a gene seen at time $s$ in an unrepresented species descend from a descendant seen at time $t<s$ in an unrepresented species.  Each of the above functions can be expressed as differential equations describing evolution backwards in time by considering the set of possible events that change the relevant probability. These can be derived analogously to \cite{tofigh_2009, Stadler:2011fk, Szollosi:2012fkk} and can be found in the Appendix. }

Given a rooted gene tree topology $G$ we can now calculate the probability $p(G | S, \mathcal{M})$ of observing $G$, where $\mathcal{M}$ denotes the parameters of the model, by summing over all possible paths along $S$ and over all complete phylogenies compatible with the species tree spanning the $n$ species of the sample. We can sum over all paths by recursively mapping the branches of $G$ onto branches of $S$ generalising the ODT models algorithm\cite{Szollosi:2012fkk} to include evolution along unrepresented species {\color{black}(cf.\ Fig.\ref{rec} and \ref{diag} in the Appendix)}. 

A branch of $G$ represents the evolution of a gene copy for which i) if the branch is nonterminal, all observed descendants descend from one of the two daughter gene lineages which emerge from the gene tree node in which the branch terminates, or ii) if the branch is terminal, a gene is observed in one of the genomes mapping to a leaf of $S$. To describe possible paths along $S$ that this gene copy may take before arriving at the gene tree node in which it terminates, we must consider five events: i) single-copy evolution along branch $e$ of $S$ described by $G_e$, ii) single-copy evolution outside $S$ described by $\bar G$; iii) speciation from a branch of $S$ to an unrepresented species such that only descendants of this copy are observed; iv) transfer such that only descendants of the transferred copy are observed and v) speciation represented in $S$ such that only one of the descending copies leaves an observed descendant. Each of these events leads to a single gene copy with observed descendants. The gene tree node in which the branch terminates can correspond to three possible events i) a duplication; a speciation represented in $S$; ii) a speciation not represented in $S$; or iii) a transfer. Each of these events leads to two gene copies with observed descendants.

To derive the recursion expressing the probability of $G$ as the sum over possible paths along $S$ we discretize time along $S$ keeping track of speciation times $t_i$ along $S$. Speciations represented in $S$ define the time intervals $[0,t_{1}),\dots,[t_i,t_{i+1}),\dots [t_{n-1},t_{n-1})$ referred to as \emph{time slices} \cite{tofigh_2009,doyon_2011} with indices $0,\dots, i, \dots n$. {\color{black} We further divide} each time slice into $D$ equal time intervals of height $\Delta t_i = (t_{i+1}-t_i) / D $. 

The probability of the gene lineage leading to node $u$ of $G$ being seen on branch $e$ of $S$ at time $t+\Delta t$ given the probabilities at time $t=t_i+ \Delta t_i$ is 
\begin{align}
 P_e (u,t+\Delta t_i) =& G_e (t+\Delta t,t ) P_e(u,t) \label{Pe}\\
& +\left\{\delta \Delta t_i \right\} P_e(v,t) P_e(w,t) \nonumber \\
& +\left\{\sigma \Delta t_i \right\} \bar P(v,t) P_e(w,t)\nonumber \\
& +\left\{\sigma \Delta t_i \right\} P_e(v,t) \bar P(w,t)\nonumber \\
& +\left\{\sigma \Delta t_i \right\} \bar P(u,t) E_e(t), \nonumber 
\end{align}
where $\bar P(u,t)$ denotes the probability of the gene lineage leading to node $u$ of $G$ being seen in an unrepresented species at time $t$, $v$ and $w$ descend from $u$ in $G$. {\color{black} As shown in Fig.\ref{diag}a in the Appendix, the terms correspond to i) no event with an observed descendent; ii) birth of two gene linages by duplication, such that both leave observed descendants; iii) and iv) birth of two gene linages with observed descendants as a result of an unrepresented speciation; and finally, v) unrepresented speciation followed by the loss of the copy in branch $e$ such that only the copy in the unrepresented phylogeny leaves an observed descendant.} In the above expression we only consider indirect transfers that involve a speciation, see the Appendix for the full expression. 

The probability of being seen in such an unrepresented species is:
\begin{align}
 \bar P(u,t+\Delta t_i) & = \bar G (t+ \Delta t_i,t ) \bar P(u,t ) \label{Pa}\\
& + \left\{ (2 \sigma + \delta + \frac{N-n_i }{N-1}\tau )\Delta t_i \right\} \bar P(v,t) \bar P(w,t) \nonumber\\
& + \sum_{e\in \mathcal{E}_i}\left\{\frac{\tau \Delta t_i }{N-1} \right\} \bar P(v,t) P_{e}(w,t) \nonumber \\ 
& + \sum_{e\in \mathcal{E}_i} \left\{\frac{\tau \Delta t_i }{N-1} \right\} P_{e}(v,t) \bar P(w,t) \nonumber \\
& + \sum_{e\in \mathcal{E}_i} \left\{\frac{\tau \Delta t_i }{N-1} \right\}\bar E(t) P_e(u,t ) \nonumber
\end{align}
where $\mathcal{E}_i(S) $ denotes the set of branches of $S$ in time slice $i$.  {\color{black} As shown in Fig.\ref{diag}b, the terms correspond to i) no event with an observed descendent; ii) birth of two gene linages by speciation, duplication or transfer, such that both leave observed descendants; iii) and iv) birth of two gene linages with observed descendants as a result of transfer back to the represented phylogeny; and finally, v)  transfer back to the represented phylogeny following which the copy in the unrepresented donor linage does not leave an observed descendant.} Terms involving {\color{black} gene lineages} $v$, $w$ are zero if $u$ is a leaf of $G$ in both the above expressions.

At speciation times $t=t_i$ where branches $f$ and $g$ descend from $e$ in $S$, a represented speciation takes place that may be followed by a loss:
\begin{align}
P_e(u,t) &= P_f(v,t) P_g(w,t) + P_f(w,t) P_g(v,t) \label{PeS}\\
 &+ P_f(u,t) E_g(t) + E_f(t) P_g(u,t).\nonumber 
\end{align} 
{\color{black} The terms (cf.\ Fig.\ref{diag}c) correspond to i) and ii) represented speciation such that both resulting gene lineages lead to observed descendants; and iii) and iv)  represented speciation such that only one of them do.} Finally at time $t=0$ on each terminal branch $e$ of $S$ the presence of observed genes is expressed as:
\begin{align}
 P_e(u,0) = 
\left\{
 \begin{array}{l l}
 1 & \text{if $u$ is a leaf of $G$ found in $e$ }\\
 0 & \text{otherwise} \\
 \end{array} \right. \label{Pe0}
\end{align}

{\color{black} As illustrated in figures \ref{rec}b and \ref{diag} each term in equations \ref{Pe}-\ref{Pe0} above corresponds to a series of speciation, duplication and transfer events that recursively draw the gene phylogeny into the species tree.} The recursion calculates the probability of a gene tree with $m$ genes in $O(D n^2 m)$ steps, as there are fewer than $n $ branches in each time slice and $n$ time slices. Summing over roots of $G$ can be accomplished with identical complexity using double recursion. The most likely reconciliation can be recovered by tracing back along the sum choosing at each step the event with the highest probability. 

Calculating the probability of a gene tree requires knowledge of the ultrametric species tree $S$, with branch lengths corresponding to time, the rate of duplication $\delta$, transfer $\tau$ and loss $\lambda$, as well as the parameters of the speciation dynamics, the species replacement rate $\sigma$ and the total number of species $N$. The number of parameters is reduced, if we assume the time to the common ancestor of the sampled species to correspond to its expected value under speciation dynamics. Choosing units such that $S$ is of unit height this corresponds to the choice $\sigma=2N$. Furthermore, under the present choice of parameters and time scale, the probability of a gene tree and its maximum likelihood reconciliation depends only very weakly on $N$, as long as the condition $n\ll N$ is satisfied. This is the case because the expected number of transfers between branches of $S$ is nearly independent of $N$. In particular if we assume that a gene lineage returns at most once to $S$ we arrive at the result derived in equation \ref{indirect_eq} according to which the number of transfers is independent of $N$.

\section*{Routes to cyanobacterial genomes}

{\color{black} To carry out a preliminary analysis of the signal for evolution outside the represented phylogeny in real data, we considered a set of $473$ single-copy gene families} present in the genome of at least $34$ of $36$ cyanobacteria and use the dated species tree reconstructed in \cite{Szollosi:2012fkk}.  {\color{black} We choose single-copy near universal gene families as they are expected to be i) relatively slowly evolving and hence to harbor a strong signal of homology and yield high quality alignments, and ii) they can be assumed to be well described by a single set of uniform duplication, transfer and loss rates, at least in contrast to more complex datasets composed of multi-copy families.} For each family, gene tree topologies and duplication, transfer and loss rates that maximize the joint likelihood\cite{Maddison:1997ly,Szollosi:2012fk} were inferred as described in the Appendix. Using these results $1000$ reconciliations per family were sampled by stochastic backtracking along the sum over reconciliations.

\begin{figure}
\begin{center}
\centerline{\includegraphics[width=1.\columnwidth]{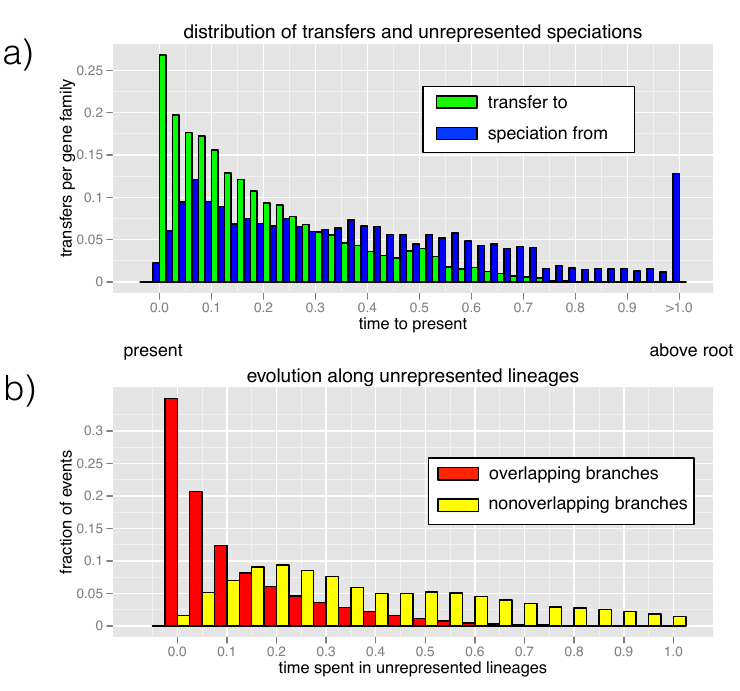}}
\caption{{\bf Lateral gene transfer events for 36 cyanobacteria.} For 473 near universal single-copy families from 36 cyanobacterial genomes gene trees that maximize the joint likelihood were reconstructed. For the trees obtained $1000$ reconciliations were sampled. a) shows the distribution of transfer events (light bars, green {\color{black}online}) and the preceding speciation events (dark bars, blue {\color{black}online}). The final bin summarizes all events occurring above the root of $S$. b) shows the distribution of the time spent by transferred genes evolving along unrepresented species for transfers between overlapping branches (dark bars, red {\color{black}online}, $72.2\%$ of transfers) and transfers between nonoverlapping branches (light bars, yellow {\color{black}online}, $27.8\%$ of all transfers). Both sets of bins sum to unity. Time units are chosen such that the height of the root of $S$ is $1.0$. The age of the root falls in the $3500-2700$ Mya interval \cite{Falcon:2010fk,Szollosi:2012fkk}. {\color{black} Data is available from Dryad under doi:10.5061/dryad.27d0g.}\label{data}}. 
\end{center}
\end{figure}

On average we found $0$ duplications, $2.15$ transfers and $2.56$ losses per family. The distribution in time of transfer events and the preceding speciations to unrepresented species are shown in Fig.\ref{data}a. The majority of transfers occur between branches of $S$ that overlap in time, hence the resulting gene tree carries no topological signature of the length of time spent evolving along unrepresented lineages. Transfers between branches n{{\bf Lateral gene transfer events for 36 cyanobacteria.} For 473 near universal single-that do not overlap in time, for which the gene tree topologies explicitly record evolution outside the represented tree, correspond to $27.8\%$ of all transfers. About a fifth of these ($5.9\%$ of all transfers) branch above the root indicating transfer from outside the sampled diversity of cyanobacteria. The median interval of time spent evolving in unrepresented lineages is $0.083$ (or $~222$ {\color{black}million years, hence forth} myr) for transfers between overlapping branches and $0.39$ (or $~1000$ myr) for transfers between nonoverlapping branches. Similar values are obtained if we consider only the maximum likelihood reconciliations, except for the median interval of time spent evolving in unrepresented lineage for transfers between overlapping branches which is only $0.0028$ (or $~8.1$ myr corresponding to the minimum length allowed by time discretization). The corresponding value for transfers between nonoverlapping branches, $0.36$ ( or $~990$ myr), is nearly identical to the value above. 

{\color{black} We emphasize that an important caveat of these results is that the accuracy of our method to infer correct reconciliations and gene topologies has not been assessed. This could be accomplished by explicit simulations of gene family evolution along the complete phylogeny. Such simulations are, however, outside the scope of the current publication, as they are technically challenging due to the large number of species in the complete phylogeny, and since they must address a potentially long list of possible questions. In lieu of simulation it is possible to examine the posterior support of individual transfer events, which, as described in the appendix, can be calculated as the fraction of times we find a given transfer event among the sampled reconciliations for each family. Using this measure, we find that transfers are well supported with $66.8\%$ of transfer events having support over $0.95$.}

It is also important to discuss to what extent we can expect {\color{black} observed transfers between nonoverlapping branches} to be robust to increasing the number of sampled species. Consider the extreme case that all $N$ extant species are sampled. It is clear that transfers between overlapping branches of $S$ (red in Fig.\ref{data}b) may correspond to transfer between nonoverlapping branches of the full phylogeny spanned by all $N$ extant species. To ascertain how often we expect the opposite to occur, to have a transfer between nonoverlapping branches of $S$ correspond to transfers between overlapping branches of the full phylogeny spanned by all $N$ extant species, we need to estimate how often we expect to sample an extant descendant of the unrepresented donor lineage involved in a transfer between nonoverlapping branches of $S$ (light bars, yellow {\color{black}online}, in Fig.\ref{data}b). Assuming a tree with unit height the total branch length of the full phylogeny under Kingman's coalescent is of the order of $\log(N)$, while the total branch lengths including extinct species is of the order $N$. Thus, we expect that only a vanishing fraction of the order $\log(N)/N$ of donor lineages have left extant descendants.{\color{black} This implies that not only do most transfers involve speciation to, and evolution along branches of the complete phylogeny, but the majority of these donor lineages have gone extinct. Consequently, most transfers between nonoverlapping branches of $S$ correspond to transfers between nonoverlapping branches of the full phylogeny where the donor lineage has gone extinct.} 

{\color{black} In summary, we find that nearly a third -- $27.8\%$ -- of transfers evolve on
average a billion years along lineages unrepresented in the phylogeny - most often, in fact, along extinct lineages, and only a moderate fraction of transfers originate from outside the cyanobacteria.} Furthermore, both of these estimates are conservative, as increasing the number of sampled species is expected to lead to an increase in the ratio of transfers between nonoverlapping branches, and to a decrease in the fraction of transfers from outside of cyanobacteria. The first of the above results, however, applies only to transfers between branches of $S$, i.e., transfers observed for the $n=36$ cyanobacteria considered. For the complete set of transfers between branches of the full phylogeny the fraction of transfers evolving along extinct linages is potentially different, e.g.\ a macroscopic fraction of transfers are expected to correspond to direct transfers between its branches.    

\section*{Discussion}

The results developed above are conditional on two crucial assumptions: i) that the number of sampled species is small compared to the total number of species, and ii) the evolution of gene lineages can be treated as independent, both in the represented and the unrepresented part of the phylogeny. As we argue above, if genes are duplicated, transferred and lost independently, the former assumption (i.e., $n\ll N$) implies that the evolution of genes outside the represented phylogeny can also be treated as independent, even if the complete phylogeny is not specified. 

We also make the assumptions that iii) transfer occurs with identical rate between any two species and iv) that the time to the last common ancestor of the sampled species corresponds to its expected value under the speciation dynamics. These conditions serve to simplify the development of the above arguments and can be relaxed without affecting our conclusion that the majority of transfers involve evolution along extinct or unsampled species. Relaxing condition iv is straightforward. Concerning assumption iii, if, for example, transfer occurs preferentially between species that are more closely related\cite{Andam:2011uq}, the scenarios shown in figure \ref{direct} are affected to an identical extent because the last common ancestor of branch $e$ and either branch $f$ (the donor lineage for dark grey paths, blue {\color{black}online}) or any extinct species that descends from an unrepresented speciation along $f$ (a donor lineage along light grey paths, red {\color{black}online}) is the same. Conversely, there are known cases, e.g. the transfer of thermostable enzymes from thermophilic archaea to thermophilic bacteria\cite{Nelson:1999ys,Nesbo:2001zr,Brochier-Armanet:2007vn}, of preferential transfer between distantly related taxa due to shared ecology. In this second case, we expect to observe genes preferentially transferred from phylogenetically distant taxa to lead to an excess of transfers descending from above the root of the sampled species for which topologically equivalent direct transfers do not exist. On a more practical ground, however, relaxing the assumption of homogeneous rates of transfer between lineages might seriously complicate the computation of the likelihood, as it would require modelling the distribution of the rates of transfers from and to unrepresented lineages.

More importantly, as long as these conditions are met, it is possible to extend the above results to more general models of speciation. Modelling variation in $N$, the total number of species, over geological times, could be of particular interest. Indeed, a corollary of the observation that LGT events record evolutionary paths along the complete species tree is that the phylogenies of genes from a limited sample of extant species carry information about extinct lineages, and therefore about the size and dynamics of ancient biodiversity. In fact, patterns of gene transfer may be even more informative about past biodiversity than the species tree itself. Drawing an analogy with population genetics, inferring biodiversity dynamics based on species trees \cite{Nee:2001fk,Morlon:2010uq,Stadler:2011fk} is similar to inferring past demography based on single-locus data. Single-locus inference is limited by the intrinsic stochasticity of Kingman's coalescent, in particular in the deep part of the genealogy. Lateral gene transfers, on the other hand, are analogous to multiple loci\cite{Heled:2008ve}, and as such, have the potential to increase the statistical power for inferring past biodiversity.

\begin{acknowledgments}
We thank B. Boussau and all the members of the Bioinformatics and Evolutionary Genomics Group for discussions of the results and comments on the manuscript. GJSz is supported by the Marie Curie Fellowship 253642 ``Geneforest''. This work was granted access to the Institut National de Physique Nucl\'eaire et de Physique des Particules' (IN2P3) computing centre. This project was supported by the French Agence Nationale de la Recherche (ANR) through Grant ANR-10-BINF-01- 01 ``Ancestrome''.
\end{acknowledgments}

\bibliographystyle{sysbio}

\newpage

\clearpage
\newpage
\appendix
\renewcommand{\thefigure}{A\arabic{figure}}
\addtocounter{figure}{-4}
\renewcommand{\theequation}{A\arabic{equation}}
\addtocounter{equation}{-6}

\begin{figure*}
\begin{center}
\centerline{\includegraphics[width=2.\columnwidth]{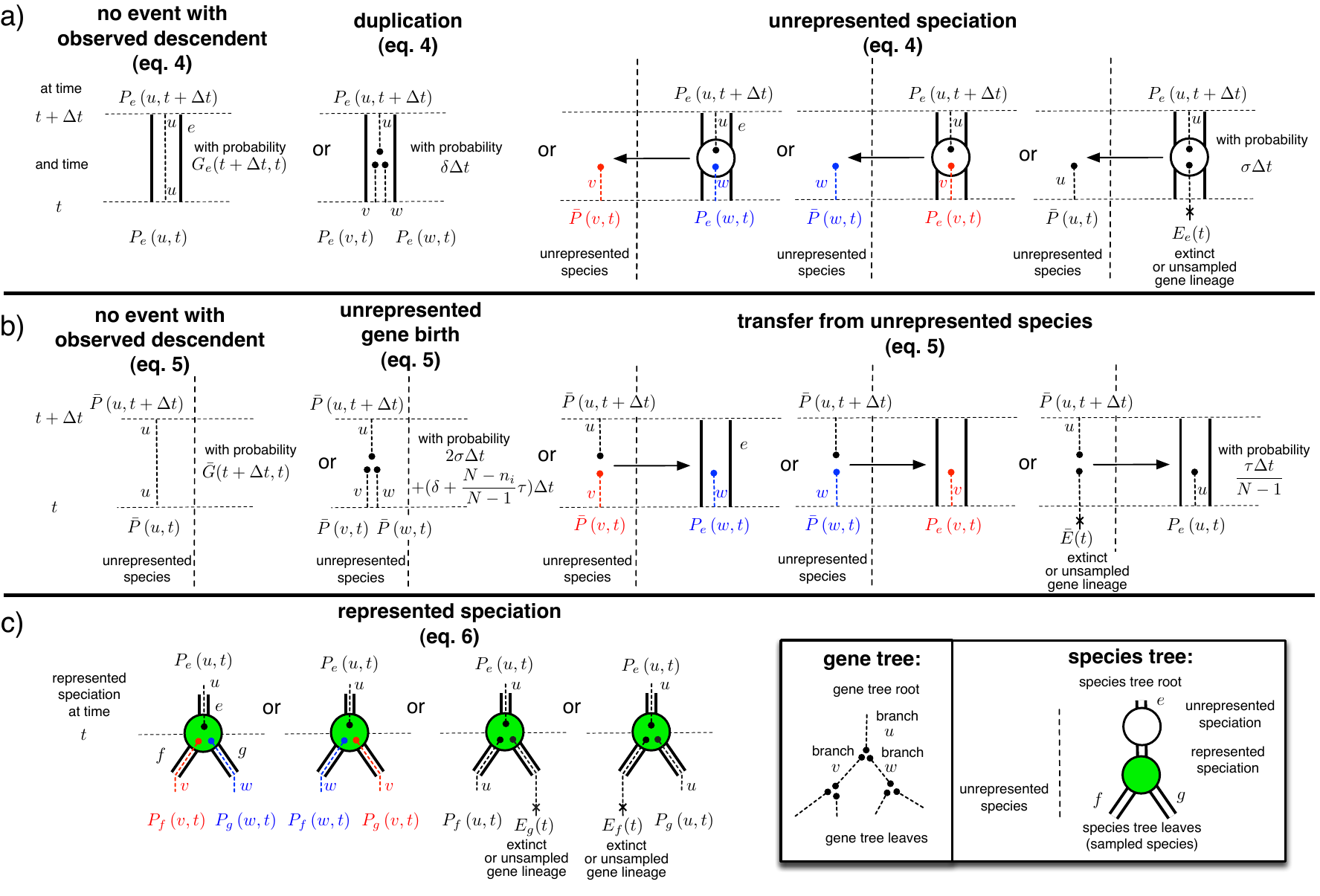}}
 \caption{{\color{black}\bf Diagrams corresponding to reconciliation events. }{\color{black} Each diagram corresponds to a term in equations \ref{Pe}-\ref{Pe0}, with diagrams following each other in the same order as terms in the indicated equation. a) depicts events that start with a gene lineage $u$ in represented branch $e$ of $S$ at time $t+\Delta t$; b) events which start with a gene lineage $u$ in an unrepresented species at time $t+\Delta t$; finally c) corresponds to represented speciation events in $S$. To illustrate the correspondence between terms and equations consider the third diagram in the top row (a) depicting an unrepresented speciation and the corresponding (third) term in equation \ref{Pe}. This term,  $P_e(u,t + \Delta t) = \cdots + \{\sigma \Delta t\}P_e(v,t) \bar P(w,t) + \cdots $, describes the probability that gene lineage $u$ seen at time $t+\Delta t$  is succeeded as a result of an unrepresented speciation by two gene linages ($v$ and $w$) one of  which ($w$) is present in the same branch $e$ as $u$ while the other ($v$) resides in an unrepresented species.} } 
\label{diag}
\end{center}
\end{figure*} 

\section*{Appendix}
{\color{black}
The dynamic programming algorithm described in equations \ref{Pe}-\ref{Pe0} calculates the likelihood of a gene tree given the species tree $S$ and the rates of speciation, duplication, transfer and loss. As illustrated in Fig.\ref{rec}, the likelihood is calculated in a piece-wise independent manner, from the evolution of gene copies that appear as single gene lineages when observed from the present. Since in contrast to Fig.\ref{rec} we do not know the exact evolutionary scenario we must sum over all reconciliations. This process can be represented as summing over all possible ways to draw the gene tree into the species tree using a the set of events shown in Fig.\ref{diag}. The diagrams in Fig.\ref{diag}, and the corresponding terms in equations \ref{Pe}-\ref{Pe0}, are expressed using two types of functions describing the evolution of single gene lineages: i) the extinction probabilities $E_e$  and $\bar E$ that give the probability of gene present on, respectively,  branch $e$ of $S$ or an unrepresented species having no descendant at time $t=0$ in the genome of any of the $n$ sampled species; ii) the single gene propagators $G_e(s,t)$ and $\bar G(s,t)$ corresponding to the probability that all sampled descendants of the gene seen at time $s$, respectively, on branch $e$ of $S$, or in an unrepresented species, descend from the gene present at a later time $t$ in the same species.

Below we provide the expressions for each of these functions that can be derived using the theory of birth-and-death processes. We also discuss the independence assumption in relation to the single gene propagators, write down the complete form of equation \ref{Pe} and describe the details of the data analysis presented in the main text.  
}

\subsection*{Evolution of single genes}
{\color{black}  The forward Kolmogorov equations describing single gene extinction and propagation can be derived analogously to  \cite{tofigh_2009, Stadler:2011fk, Szollosi:2012fkk}. The main differences is that here we also consider the speciation dynamics.}

The extinction probability for branch $e$ of $S$:
\begin{align}
 \frac{\mathrm{d}}{\mathrm{d} t} E_e = & + \left\{ \lambda \right\} (1-E_e) \label{ee}\\
& -\left\{ \delta (1-E_e) \right\} E_e \nonumber \\ 
& - \left\{ \sum_{f \in \mathcal{E}_i(S) } \frac{\tau}{N-1} (1-E_f) \right\} E_e \nonumber\\
& -\left\{ (\sigma + \frac{N-n_i}{N-1}\tau )(1-\bar E) \right\} E_e \nonumber \\
E_e(t_e^{\mathrm{end}}) = &\left\{ 
 \begin{array}{l l}
 0 & \text{if $t_e^{\mathrm{end}}=0$, }\\
 E_f(t^{\mathrm{begin}}_{g}) E_g(t^{\mathrm{begin}}_f) & \text{otherwise}\nonumber \\
 \end{array} \right.,\nonumber
\end{align}
where $\mathcal{E}_i(S) $ denotes the set of branches of $S$ in time slice $i$, and $n_i$ their number. The terms correspond to i) loss ii) the rate duplications and {\color{black}iii)} transfers to represented hosts, both conditional on survival, and finally {\color{black}iv)} the rate of unrepresented speciations and transfers to unrepresented hosts, again conditional on survival. The initial conditions specify that at the end of branch $e$ the probability of extinction is $0$ if we are at time $t=0$ i.e., $e$ is a terminal branch of $S$, and the product of the extinction probability of the descendants of $e$ in $S$ otherwise. 

The extinction probability in an unrepresented species:
\begin{align}
\frac{\mathrm{d}}{\mathrm{d} t} {\bar E} = &+\left\{\sigma + \lambda \right\} (1-\bar E ) \label{bare}\\
&- \left\{ ( \sigma + \delta + \frac{N-n_i}{N-1}\tau ) (1-\bar E ) \right\} \bar E \nonumber\\
& - \left\{ \sum_{f \in \mathcal{E}_i(S) } \frac{\tau}{N-1} (1-E_f) \right\}\bar E \nonumber\\
 \bar E(0)=&1,\nonumber 
 \end{align}
Note that the term corresponding to transfer back to $S$ acts as an inhomogeneity in the absence of which the only solution is $\bar E (t)=1$. 

The single observed lineage propagator along branch $e$ of $S$:
\begin{align}
 \frac{\mathrm{d}}{\mathrm{d} t} G_e = & -\left\{ \lambda + \delta (1- 2 E_e) \right\} G_e \label{ge}\\
& - \left\{ \sum_{f \in \mathcal{E}_i(S) } \frac{\tau}{N-1} (1-E_f) \right\} G_e \nonumber\\
&-\left\{ \frac{N-n_i}{N-1}\tau (1-\bar E)\right\} G_e \nonumber\\
& - \left\{ \sigma(1-\bar E) \right\} G_e \nonumber\\
 G_e(t,t) = 1.\nonumber
\end{align}

The single observed lineage propagator in an unrepresented species:
\begin{align}
 \frac{\mathrm{d}}{\mathrm{d} t} \bar G = & - \left\{ \sigma + \lambda \right\} \bar G \label{barg}\\ 
& - \left\{ (\sigma + \delta + \frac{N-n_i}{N-1}\tau) (1-2\bar E )\right\} \bar G \nonumber\\ 
& - \left\{ \sum_{f \in \mathcal{E}_i(S) } \frac{\tau}{N-1} (1-E_f) \right\} \bar G \nonumber \\ 
\bar G(t,t) = & 1.\nonumber 
\end{align}
Note that if we set $\bar E(t)= 1$ and neglect gene birth and death, which is much slower than the speciation dynamics - i.e., $\delta+ \tau+\lambda \ll \sigma$, we recover the exponential probability of coalescence with the represented tree assumed in equation \ref{indirect_eq}. 

The propagator describing the evolution of $k$ gene copies can be expressed using the single gene copy propagator. Consider the expression for $\bar G_k(s,t)$, the probability that $k$ genes seen at time $s$ in an unrepresented species all leave a single descendant descending from the copy seen at time $t<s$ in an unrepresented species:
\begin{align}
 \frac{\mathrm{d}}{\mathrm{d} t} \bar G_k = & - \left\{ \sigma + k \lambda \right\} \bar G_k \label{bargk}
\\ 
& - \left\{ \sigma (1-2\bar E_k ) + k ( \delta + \frac{N-n_i}{N-1}\tau) (1-2\bar E )\right\} \bar G_k \nonumber\\ 
& - \left\{ \sum_{f \in \mathcal{E}_i(S) } \frac{k \tau}{N-1} (1-E_f) \right\} \bar G_k \nonumber \\ 
\bar G(t,t)_k = & 1.\nonumber 
\end{align}
Since \ref{indep} implies $ - \sigma \bar G_k -\sigma (1-2\bar E_k ) \bar G_k \simeq \sigma 2 (1- \{ \bar E\}^k ) \bar G_k$ and neglecting second and higher order terms in $1-\bar E$ gives $2 (1- \{ \bar E\}^k ) \simeq 2 k (1- \bar E ) $ and the above can be written as
\begin{align}
 \frac{\mathrm{d}}{\mathrm{d} t} \bar G_k \simeq & - k \left\{ \sigma + \lambda \right\} \bar G_k \label{bargk2}\\ 
& - k \left\{ \sigma (1-2\bar E ) + ( \delta + \frac{N-n_i}{N-1}\tau) (1-2\bar E )\right\} \bar G_k \nonumber\\ 
& - k \left\{ \sum_{f \in \mathcal{E}_i(S) } \frac{ \tau}{N-1} (1-E_f) \right\} \bar G_k, \nonumber 
\end{align}
 which has the solution $\bar G_k = \{\bar G \}^k$. Analogous reasoning can be used to show that $E_e(t)$ and $G_e(s,t)$ can be used to factor the respective functions describing the evolution of multiple gene copies.

\subsection*{The probability of a gene tree}

The expressions for $P_e(u,t)$ and $\bar P(u,t)$ are derived under the approximation that unrepresented speciation and extinction is independent per gene, as discussed above. The full expression for $P_e (u,t+\Delta t_i)$ including terms corresponding to direct transfers and indirect transfers that depart $S$ via a preceding transfer events, which are neglected in equation \ref{Pe}, is:
\begin{align}
 P_e (u,t+\Delta t_i) =& G_e (t+\Delta t,t ) P_e(u,t) \\
& +\left\{\delta \Delta t_i \right\} P_e(v,t) P_e(w,t) \nonumber \\
& +\sum_{f\in \mathcal{E}_i} \left\{\frac{\tau \Delta t_i }{N-1}\right\} P_e(v,t) P_f(w,t)\nonumber \\
& +\sum_{f\in \mathcal{E}_i} \left\{\frac{\tau \Delta t_i}{N-1} \right\} P_f(v,t) P_e(w,t)\nonumber \\
& +\sum_{f\in \mathcal{E}_i} \left\{\frac{\tau \Delta t_i }{N-1}\right\} P_f(u,t) E_e(t)\nonumber \\
& +\left\{(\sigma + \frac{N-n_i}{N-1}\tau)\Delta t_i \right\} \bar P(v,t) P_e(w,t)\nonumber \\
& +\left\{(\sigma + \frac{N-n_i}{N-1}\tau)\Delta t_i \right\} P_e(v,t) \bar P(v,t)\nonumber \\
& +\left\{(\sigma + \frac{N-n_i}{N-1}\tau)\Delta t_i \right\} E_e(t) \bar P(u,t), \nonumber 
\end{align}

\subsection*{Routes to cyanobacterial genomes}

The dataset was constructed for near universal single-copy genes from all $36$ cyanobacterial genomes found in version 5 the HOGENOM database\cite{Penel:2009ly}. Amino acid sequences were extracted for each family that had a single-copy in at least $34$ of the $36$ cyanobacterial genomes. For each family sequences were aligned using MUSCLE \cite{Edgar:2004kx} with default parameters. The multiple alignment was subsequently cleaned using GBLOCKS \cite{Talavera:2007vn} with the options:
\begin{equation}
\text{``{\tt -t=p -b1 50 -b2 50 -b5=a -t=p}''.}\label{gblocks}
\end{equation}
Subsequently we inferred a gene topology $G$ that maximizes the joint likelihood
\begin{align}
 p_{\mathrm{exODT} }(G | S,\delta,\tau,\lambda,\sigma,N ) \times p_{\mathrm{Felsenstein} }(\mathrm{alignment} | G),
\end{align}
where the first term corresponds to the likelihood of observing the unrooted gene tree topology $G$ according to the exODT model developed above (equations \ref{Pe} and \ref{Pa}), while the second term corresponds to the classic Felsenstein likelihood\cite{felsenstein81} of the alignment. For the exODT model we fixed the parameter values $N=10^6$ and $\sigma=2N$, used the dated phylogeny from \cite{Szollosi:2012fkk} and estimated global gene birth and death rates as described below. To calculate the Felsenstein likelihood we used the Bio++ library \cite{Dutheil:2006uq} with an LG+$\Gamma$4+I model. {\color{black} Alignments and reconstructed gene trees are available from Dryad under doi:10.5061/dryad.27d0g.}

Gene trees inference was performed in a two step approach: 
\begin{description}
\item[Initial estimate]\hskip 0.75\columnwidth
\begin{enumerate}
\item using the DTL rates $\delta=1.0\times10^{-2},\ \tau=1.0\times10^{-2}, \lambda=2.0\times10^{-2}$ for each family the joint likelihood was calculated for all nearest neighbor interchanges \cite{Felsenstein:2004dq} (NNIs) and a move was accepted if it improved the joint likelihood. 
\item for the set of trees obtained global DTL parameters were estimated that maximize the product of the joint likelihood of all $473$ gene families.
\end{enumerate}
\item[Final estimate]\hskip 0.75\columnwidth
\begin{enumerate}
\item using the obtained DTL rates for each family the joint likelihood was calculated for all nearest neighbor interchanges \cite{Felsenstein:2004dq} (NNIs) and a move was accepted if it improved the joint likelihood. 
\item for the set of trees obtained global DTL parameters were again estimated with the results: $\delta=1.010\times10^{-5},\ \tau=4.438\times 10^{-3}, \lambda = 1.015 \times 10^{-1}$.
\end{enumerate}
\end{description}
Before performing NNIs starting gene tree topologies were estimated using an amalgamation approach \cite{David:2011zr} wherein the Felsenstein likelihood was approximated using conditional clade probabilities\cite{Hohna:2012qf} based on posterior sample of 10000 tree topologies obtained using PhyloBayes using an LG+$\Gamma$4+I substitution model. 

{\color{black} The support of transfer events was measured based on a posterior sample of $1000$ reconciliations per family. For each family we assessed the support of all transfer events in the reconciliation that was seen the largest number of times. Two transfers were considered equivalent if they involved the transfer of the same gene linage between identical branches of the species tree. }

\end{document}